\documentclass[structabstract]{aa}

\usepackage{graphicx}
\usepackage{txfonts}

\usepackage[authoryear,round]{natbib}

\begin{document}
   \title{Compression Behaviour of Porous Dust Agglomerates}

   %\subtitle{Compressibility}

   \author{A. Seizinger,
          \inst{1}
          R. Speith,
          \inst{2}
          \and
          W. Kley\inst{1}
          }

   \institute{Institut f\"ur Astronomie and Astrophysik, Eberhard Karls Universit\"at T\"ubingen,\\
              Auf der Morgenstelle 10c, D-72076 T\"ubingen, Germany\\
              \email{alexs@tat.physik.uni-tuebingen.de}
              \and
              Physikalisches Institut, Eberhard Karls Universit\"at T\"ubingen,\\
              Auf der Morgenstelle 14, D-72076 T\"ubingen, Germany\\
             }

   \date{Received 20.01.2012; accepted 11.03.2012}
 
  \abstract
   {The early planetesimal growth proceeds through a sequence of sticking collisions of dust agglomerates. Very uncertain is still
  the relative velocity regime in which growth rather than destruction can take place. The outcome of a collision depends on the bulk properties of the porous
  dust agglomerates.}
   {Continuum models of dust agglomerates require a set of material parameters that are often difficult to obtain from laboratory experiments. 
   Here, we aim at determining those parameters from ab-initio molecular dynamics simulations. Our goal is to improve on the existing model
   that describe the interaction of individual monomers.}
   {We use a molecular dynamics approach featuring a detailed micro-physical model of the interaction of spherical grains.
     The model includes normal forces, rolling, twisting and sliding between the dust grains. We present a new treatment of wall-particle
    interaction that allows us to perform customized simulations that directly correspond to laboratory experiments.}
   {We find that the existing interaction model by Dominik \& Tielens leads to a too soft compressive strength behavior for uni- and omni-directional
   compression. Upon making the rolling and sliding coefficients stiffer we find excellent agreement in both cases. 
   Additionally, we find that the compressive strength curve depends on the velocity with which the sample is compressed.}
   {The modified interaction strengths between two individual dust grains will lead to a different behaviour of the whole
    dust agglomerate. This will influences the sticking probabilities and hence the growth of planetesimals. The new parameter set might possibly
    lead to an enhanced sticking as more energy can be stored in the system before breakup.}


   \authorrunning{Seizinger et al.}
   \maketitle
%________________________________________________________________
\section{Introduction}
Unraveling the question of planetesimal formation is a crucial issue for the core accretion model of planet formation proposed by \citet[][]{Pollack:1996}. To this day it remains unclear how dust and ice particles can grow several orders in magnitude from micron to kilometer sized objects. In the beginning micron sized dust grains (further referred to also as monomers or particles) may grow by low velocity, hit-and-stick collisions resulting in highly porous fractal aggregates \citep{Kempf:1999, Blum:2000b}. As the aggregates grow larger their motion increasingly decouples from the surrounding gas leading to higher collision velocities \citep[][]{Weidenschilling:1977}. With increasing impact energy colliding aggregates get compacted \citep{Blum:2000a, Suyama:2008, Wada:2008, Paszun:2009}. At some point relative velocities become large enough that fragmentation is supposed to set in \citep[e.g.\,][]{Blum:2008}, which may limit the collisional growth of aggregates. Apart from fragmentation, radial drift and bouncing \citep[e.g.\,][]{Langkowski:2008, Weidling:2009, Guettler:2010} may hamper the growth of planetesimals \citep{Zsom:2010}. Despite these obstacles, experimental evidence indicates possible growth in high speed collisions (a few $10\,\mathrm{m/s}$) through sticking and reaccretion of ejecta \citep{2005Icar..178..253W, 2009MNRAS.393.1584T} or the sweep-up of smaller particles \citep{2012arXiv1201.4282W}. To understand the possible growth regimes requires material properties and simulations beyond the current data base.

In the context of planetesimal formation, collisions of micron sized aggregates have been studied theoretically using a molecular dynamics approach \citep[e.g.][]{Dominik:1997, Wada:2007}. Here, the motions of all grains that make up the aggregate are followed individually, considering suitable interaction forces. However, billions of such grains would be necessary to model meter-sized objects. For computational reasons it is therefore necessary in such cases to use continuum models to simulate collisions between macroscopic aggregates. Smooth Particle Hydrodynamics (SPH) constitutes such an approach \citep{Sirono:2004, Schaefer:2007, Geretshauser:2010}. However, the outcome of such simulations depends strongly on the underlying porosity model \citep{Guettler:2009}. To simulate collisions of large dust boulders with SPH, several parameters describing the behaviour of the material such as the compressive-, tensile-, and shear-strength must be known in advance. 

To this day, only few attempts have been made to measure the required material parameter of porous dust aggregates in laboratory experiments. \citet{Blum:2004} studied the case of unidirectional compression, where a sample aggregate is compacted between two plates. While the bottom plate was fixed a certain load was applied to the top plate. To calculate the pressure the cross-section of the compressed aggregate had to be determined. Their samples were produced by random ballistic deposition (RBD) and featured an initial filling factor of $0.15$. The maximum filling factor they could reach was limited to approximately $0.33$ as grains began to flow outwards as the pressure in the center increased. 

Later on, \citet{Guettler:2009} used a similar approach to study the omnidirectional compression.  Their sample was put into a solid box and the load was applied by a movable piston on the top (see Fig.\,\ref{fig:cpr_exp}). As the grains could not escape the box they reached a significantly higher filling factor of roughly $0.58$ upon compression. One advantage of this approach lies in the elimination of any uncertainty in the determination of the filling factor since the volume currently occupied by the sample is unambiguous. %One might also argue that this setup reflects more closely the compression due to a collision of two macroscopic bodies.

The compressive strength of porous dust aggregates was determined numerically by \citet{Paszun:2008} using a molecular dynamics approach based on an interaction model by \citet{Dominik:1995,Dominik:1996}. Yet, they only modeled the unidirectional compression and their simulations were limited to very few particles ($\approx$ 300). Using similar material and model parameters the model has been applied to explore the growth regimes of dust agglomerates under protoplanetary conditions \citep{Suyama:2008,2009ApJ...702.1490W,2011ApJ...737...36W,2009A&A...502..845O}.

In this work we step back again and present a method to obtain the continuum material parameter of porous particle agglomerates, in particular the compressive strength curve, from ab-initio simulations using a molecular dynamics approach. Our approach is based on the model by \citet{Dominik:1995,Dominik:1996} using extensions by \citet{Wada:2007}. To test the applicability of the model we perform customized simulations of both, omni- and unidirectional compression, with a much greater number of particles of the order of $10^4$. As we will see, modifications of the model are required to properly reproduce the experimental results from \citet{Guettler:2009}. Having calibrated our model to the case of the slow compression as measured by \citet{Guettler:2009} we will subsequently present new results on how the compressive behaviour changes with increasing compression speed.

First, we briefly summarize the underlying physical model and present our extensions in Sect.\,2. Our numerical setup is explained in Sect.\,3. In Sect.\,5, we first describe the calibration of our model. Afterwards, we present our results of studying the dynamic compression and provide simple analytical approximations describing the dependence of the compressive strength on the compression speed.

%__________________________________________________________________
\section{Interaction model}
\label{sec:interaction_model}
The agglomerates used in our simulations are composed of spherical monomers of equal size. To describe the interaction between individual monomers we adopt the physical model that has been presented by \citet{Dominik:1997}. In this model elastic grains may establish adhesive contacts when touching each other. Upon deformation of these contacts kinetic energy is dissipated. The forces and torques acting upon the monomers can be derived from corresponding potentials \citep{Wada:2007}.

\subsection{Particle-Particle interaction}
\begin{figure*}
\resizebox{0.48\hsize}{!}{\includegraphics{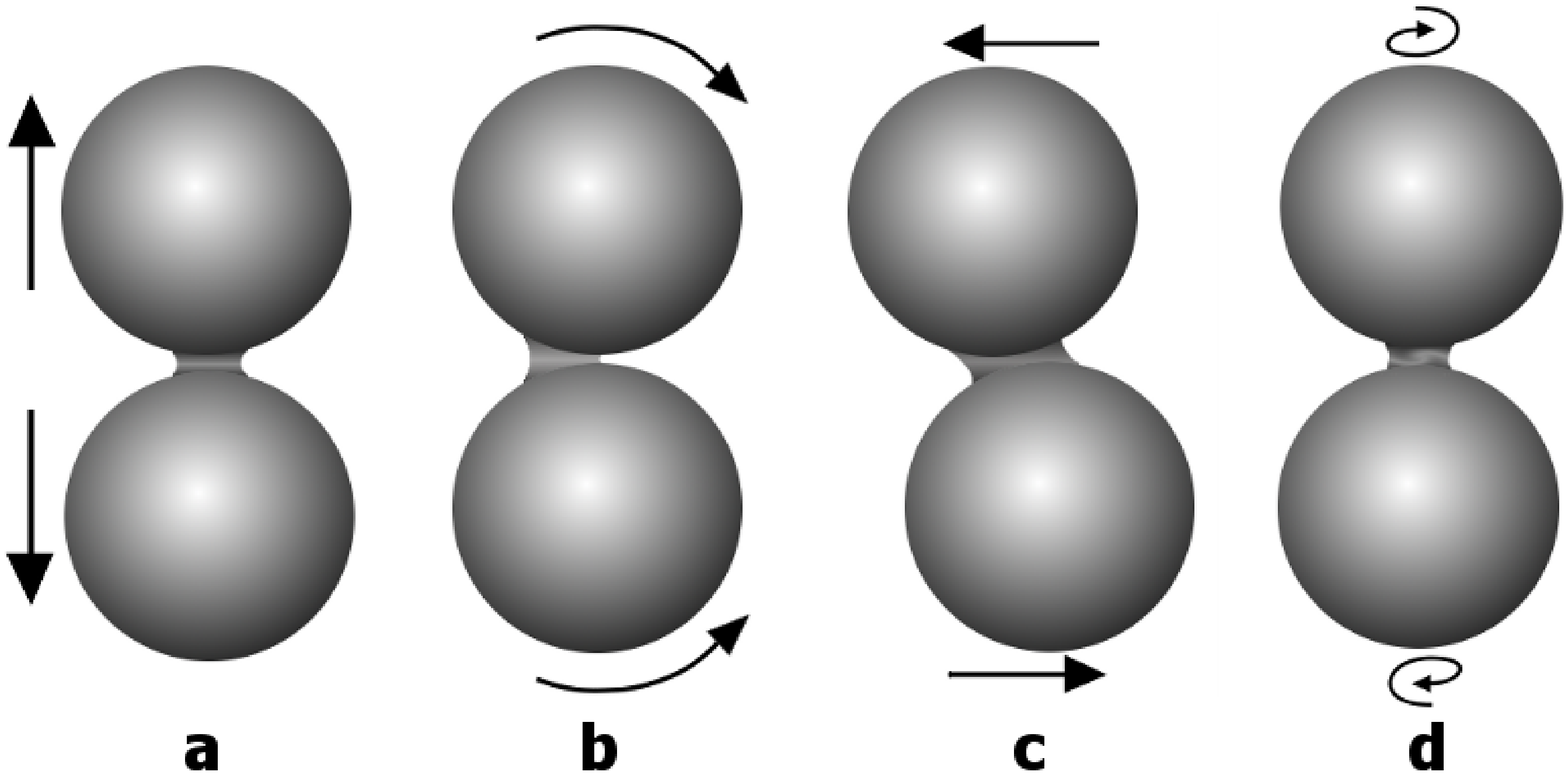}} \hfill \resizebox{0.48\hsize}{!}{\includegraphics{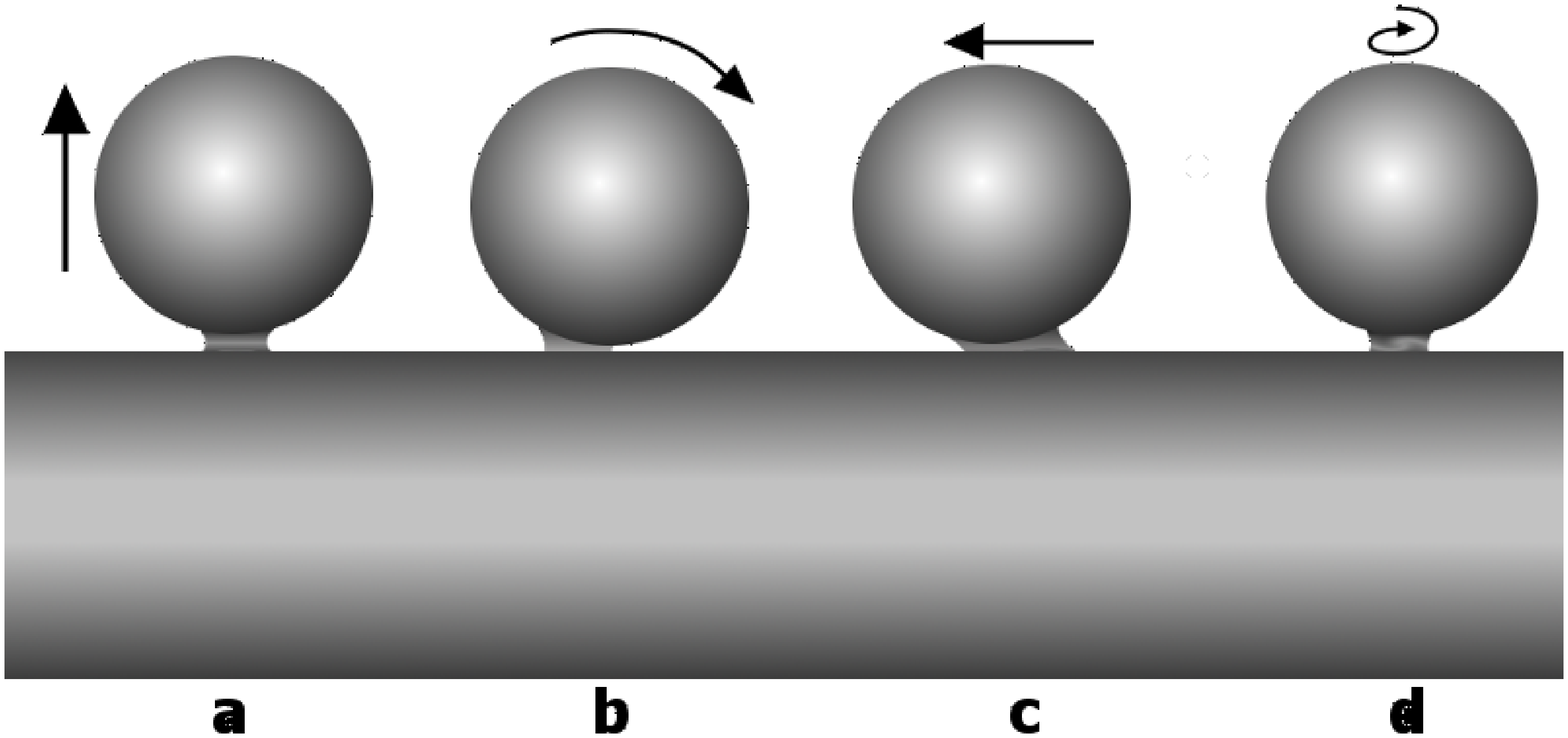}}
\caption{The four types of particle interaction: Compression/Adhesion (a), Rolling (b), Sliding (c), and Twisting (d). The left panel depicts particle-particle
 interactions while on the right the corresponding particle-wall version is displayed.}
\label{fig:interaction}
\end{figure*}

In \citet{Dominik:1997} the interaction between two spherical particles is divided into four types (see Fig.\,\ref{fig:interaction}):
\begin{enumerate}
 \item Compression/Adhesion 
 \item Rolling
 \item Sliding
 \item Twisting
\end{enumerate}
In accordance with \citet{Wada:2007} the following notation will be used in this work: $r_i$ denotes the radius of a monomer $i$, $\gamma_i$ the surface energy per area, $E_i$ the Young's modulus, $\nu_i$ the Poisson's ratio, and $G_i$ the shear modulus. Furthermore, the reduced radius $R$ is given by
\[R_{i,j} = \frac{r_i r_j}{r_i + r_j}\]
and
\begin{eqnarray*}
E^\star_{i,j} & = & \left(\frac{1-\nu_i^2}{E_i} + \frac{1-\nu_j^2}{E_j} \right)^{-1}\ ,\\
G^\star_{i,j} & = & \left(\frac{2-\nu_i^2}{G_i} + \frac{2-\nu_j^2}{G_j} \right)^{-1}\ .
\end{eqnarray*}
Note that for simplification all monomers in our simulations feature the same properties.

%material parameters
\begin{table}
 \caption[]{Material Parameters.}
 \label{tab:material_parameters} 
 \centering
 \renewcommand\arraystretch{1.2}
 \begin{tabular}{ll}
   \hline
   \noalign{\smallskip}
   Physical property & Silicate\\
   \noalign{\smallskip}
   \hline
   \noalign{\smallskip}
   Particle Radius $r$ (in $\mathrm{\mu m}$) & $0.6$\\
   Density $\rho$ (in g\, cm$^{-3}$) & $2.65$\\
   Surface Energy $\gamma$ (in mJ\, m$^{-2}$) & $20$\\
   Young's Modulus $E$ (in GPa)         & $54$\\
   Poisson Number $\nu$               & $0.17$\\
   Critical Rolling Length $\xi_\mathrm{crit}$ (in nm) & $2$\\
   \noalign{\smallskip}
   \hline
 \end{tabular}
\end{table}

\subsubsection{Material Parameters}
The basis of our simulations are the material parameters of silicate as summarized in Tab.\,\ref{tab:material_parameters}. With the exception of the surface energy $\gamma$ these values comply with \citet{Paszun:2008}, where they used $\gamma = 25\,\mathrm{mJ/m^2}$. These parameters are also in reasonable agreement with experimental data as quoted by \citet{Blum:2004,Guettler:2009}. In this work we use a slightly lower value of $\gamma = 20\,\mathrm{mJ/m^2}$ which agrees with recent measurements of \citet{Gundlach:2011}.

\subsubsection{Compression/Adhesion}
The particle interaction in the normal direction has been developed by \citet{Johnson:1971} (often referred to as JKR theory). They extend the Hertzian theory by taking adhesion due to surface forces into account.

To very briefly summarize their model let us first note that the compression length $\delta$ of two monomers, which are in contact with each other and located at $\vec{x}_1$ and $\vec{x}_2$, is defined by
\begin{eqnarray}\delta = r_1 + r_2 - \| \vec{x}_1 - \vec{x}_2 \|\ , \end{eqnarray}
where $\|\vec{u}\|$ denotes the norm of the vector $\vec{u}$.
The radius $a$ of the corresponding contact area can be obtained by
\begin{eqnarray}\frac{\delta}{\delta_0} = 3 \left(\frac{a}{a_\mathrm{0}}\right)^2 - 2 \left(\frac{a}{a_\mathrm{0}}\right)^{1/2}\ , \end{eqnarray}
where $\delta_\mathrm{0}$ and $a_\mathrm{0}$ denote the equilibrium compression length and contact radius, respectively, where the repulsive and attractive normal forces are equal.
It applies
\begin{eqnarray*}
\delta_\mathrm{0} & = & \frac{a_\mathrm{0}^2}{3 R}\ ,\\
a_\mathrm{0} & = & \left(\frac{9 \pi \gamma R^2}{E^\star} \right)^{1/3}\ .\\
\end{eqnarray*}
The force acting upon the monomers is
\begin{eqnarray}F = 4 F_\mathrm{c} \, \left[ \left(\frac{a}{a_\mathrm{0}}\right)^3 - \left(\frac{a}{a_\mathrm{0}}\right)^{3/2} \right]\ ,\label{F_normal}\end{eqnarray}
where $F_\mathrm{c} = 3 \pi \gamma R$ is the force needed to break the contact.

A new contact is established if two freely moving particles touch each other, which means $\delta = 0$. However, once a contact has formed it will not break until the compression length exceeds a certain threshold $\delta_\mathrm{c} = \left(9 / 16\right)^{1/3} \delta_\mathrm{0}$. Thus, contacts may be stretched out a little bit before they finally break.

\subsubsection{Sticking velocity}
\label{sec:sticking_velocity}
On contact formation or breaking there is a jump in the potential corresponding to the JKR force \citep[see][]{Wada:2007}, which reflects the dissipation of kinetic energy upon contact formation or breaking. \citet{Chokshi:1993} proposed that this energy dissipation is caused by the excitation of elastic waves. From this energy dissipation one can calculate the maximum velocity $v_\mathrm{stick}$ at which particles stick on head-on impacts in JKR-theory
\begin{eqnarray}v_\mathrm{stick} = 1.07 \frac{\gamma^{5/6}}{E^{\star 1/3} R^{5/6} \rho^{1/2}}.\label{eqn:v_stick}\end{eqnarray}
For micron-sized SiO$_2$ grains, we obtain velocities of the order of $0.1\,\mathrm{m/s}$. However, laboratory experiments by \citet{Poppe:2000} yield a significantly higher sticking velocity of roughly $1.2 \mathrm{m s^{-1}}$ for similar sized particles. To overcome this discrepancy, \citet{Paszun:2008} introduced another damping mechanism that dissipated additional kinetic energy upon the contact creation and thus increased the sticking velocity to match the laboratory experiments.

Since the sticking velocity has to be increased by approximately one order of magnitude a significant amount of kinetic energy has to be dissipated aside from JKR-theory. Dissipating the kinetic impact energy all at once leads to a significant change of the relative velocity between the colliding particles. As they may be in contact with other particles as well, significantly modifying the velocity of one particle during one integration step may introduce numerical hazards. For low collision velocities \citep[the compression velocity used by][was $0.05\,\mathrm{m s^{-1}}$]{Paszun:2008} the additional damping is low and therefore this problem does not arise. However, since in this work we also want to study the dynamic compression behaviour at high velocities we choose a different approach to adjust the sticking velocity as explained in the next section.

\subsubsection{Normal oscillations}
\label{sec:normal_damping}
The normal force may be both attractive or repulsive depending on the current compression length $\delta$. When two particles come too close to each other they are repelled, whereas they get attracted to each other due to adhesive surface forces while the contact is stretched out. This leads to oscillations of two monomers in the normal direction of a contact (from now on referred to as normal oscillations). In reality, one would expect that these oscillations are eventually damped away. For instance \citet{Brilliantov:2007} proposed a viscoelastic damping mechanism.

From a numerical point of view these normal oscillations constitute a major nuisance. Not only may they artificially heat up aggregates \citep{Paszun:2008} but most importantly they need to be properly resolved in time. For micron-sized SiO$_2$ grains the typical timescale of these oscillations is of the order of $10\,\mathrm{ns}$. Therefore our integration timestep is limited to $\approx 0.1-0.3\,\mathrm{ns}$. 

\citet{Paszun:2008} tackle the problem of artificial heating by introducing an additional, weak damping mechanism which has hardly any influence on the sticking velocity but eventually damps away normal oscillations. In this work we follow a similar approach. As before we consider two monomers located at $\vec{x}_1$ and $\vec{x}_2$ that are in contact with each other. The contact normal $\vec{n}_\mathrm{c}$ is 
\begin{eqnarray}\vec{n}_\mathrm{c} = \frac{\vec{x}_1 - \vec{x}_2}{\|\vec{x}_1 - \vec{x}_2\|}\ .\end{eqnarray}
The relative velocity $v_\mathrm{rel}$ in normal direction of the contact is then given by
\begin{eqnarray}v_\mathrm{rel} = \left(\vec{v}_1 - \vec{v}_2\right) \cdot \vec{n}_\mathrm{c}\ ,\end{eqnarray}
where $\vec{v}_1$, $\vec{v}_2$ denote the velocity of particle $1$ and $2$, respectively.
The viscous damping force $\vec{F}_\mathrm{damp}$ is 
\begin{eqnarray}\vec{F}_\mathrm{damp} = - \kappa v_\mathrm{rel} \vec{n}_\mathrm{c}\ ,\end{eqnarray}
where $\kappa$ is a damping constant determining the strength of the damping. We use $\kappa = 1 \times 10^{-6}\,\mathrm{g/s}$.

\subsubsection{Rolling, sliding, and twisting}
The forces and torques resulting from the tangential motion of the contact area have been formulated by \citet{Dominik:1995, Dominik:1996}. So called ``contact pointers'' \citep{Dominik:2002} provide a convenient way to track the evolution of the contact area. They are unit vectors that initially point from the center of the particle to the center of the contact area. Due to the rotation of the particles their orientation changes over time \citep{Dominik:2002}. Contact pointers are used to define the rolling, sliding, and twisting displacement which quantify how much rolling, sliding, or twisting of a contact has occurred.

Using these displacements, \citet{Wada:2007} derived the forces and torques which agree with \citet{Dominik:1995, Dominik:1996}, from corresponding potentials. All three types of interaction remain elastically as long as the displacements do not exceed certain thresholds (from now on referred to as critical displacements).

Let $\vec{n}_1$ and $\vec{n}_2$ be the corresponding contact pointers describing the contact. The rolling displacement $\vec{\xi}$, sliding displacement $\vec{\zeta}$, and twisting displacement $\vec{\phi}$ are then defined by:
\begin{eqnarray}
 \vec{\xi} & = & R\ (\vec{n}_1 + \vec{n}_2)\ , \label{eqn:roll_displ}\\
 \vec{\zeta } & = & r_1 \vec{n}_1 - r_2 \vec{n}_2 - \left(r_1 \vec{n}_1 \cdot \vec{n}_\mathrm{c} - r_2 \vec{n}_2 \cdot \vec{n}_\mathrm{c} \right) \vec{n}_\mathrm{c}\ , \label{eqn:sliding_displacement}\\
 \vec{\phi} & = & \vec{n}_\mathrm{c}(t) \int_{t_0}^t \left( \vec{\omega}_1(t') - \vec{\omega}_2(t') \right) \cdot \vec{n}_\mathrm{c}(t') dt'\ , \label{eqn:twist_displ}
\end{eqnarray}
where $t_0$ denotes the time when the contact has been established.

For the rolling motion, the forces $\vec{F}_\mathrm{r}$ and torques $\vec{M}_\mathrm{r}$ acting upon particle $1$ due to being in contact with particle $2$ are
\begin{eqnarray}
 \vec{F}_\mathrm{r} & = & 0\ ,\\
 \vec{M}_\mathrm{r} & = & - R k_\mathrm{r} \vec{n}_1 \times \vec{\xi}\ .
\end{eqnarray}
For the sliding interaction, the forces and toques are given by
\begin{eqnarray}
 \vec{F}_\mathrm{s} & = & - k_s \vec{\zeta} \frac{ \left(  r_2 \vec{n}_2 - r_1 \vec{n}_1 \right) \cdot \vec{n}_\mathrm{c}}{\|\vec{x}_1 - \vec{x}_2\|}\ ,\\
 \vec{M}_\mathrm{s} & = & - r_1 k_s \vec{n}_1 \times \vec{\zeta}\ .
\end{eqnarray}
And for the twisting interaction it applies
\begin{eqnarray}
 \vec{F}_{\mathrm{t}} & = & 0\ ,\\
 \vec{M}_{\mathrm{t}} & = & - k_t \vec{\phi}\ .
\end{eqnarray}
The constants $k_\mathrm{r}$, $k_\mathrm{s}$, and $k_\mathrm{t}$ are given by
\begin{eqnarray}
k_\mathrm{r} & = & \frac{4 F_\mathrm{c}}{R}\ , \label{eqn:k_r}\\
k_\mathrm{s} & = & 8 a_\mathrm{0} G^\star\ , \label{eqn:k_s}\\
k_\mathrm{t} & = & \frac{16}{3} G a_\mathrm{0}^3\ . \label{eqn:k_t}
\end{eqnarray}

\subsubsection{Inelastic interaction}
As already mentioned before, inelastic motion sets in when the displacements exceed a critical displacement $\xi_\mathrm{crit}$, $\zeta_\mathrm{crit}$, or $\phi_\mathrm{crit}$. Physically this corresponds to energy being dissipated upon relocation of the contact area. The critical sliding and twisting displacements have been derived theoretically \citep{Dominik:1996} as
\begin{eqnarray*}
 \zeta_\mathrm{crit} & = & \frac{2 - \nu}{16 \pi} a_0,\\ 
 \phi_\mathrm{crit} & = & \frac{1}{16 \pi}.
\end{eqnarray*}
The value of the critical rolling displacement $\xi_\mathrm{crit}$ is still debated. At first, \citet{Dominik:1997} assumed $\xi_\mathrm{crit}$ close to inner-atomic distances and chose $\xi_\mathrm{crit} = 0.2 \mathrm{nm}$. However, subsequent laboratory experiments by \citet{Heim:1999} indicate a much higher value of $\xi_\mathrm{crit} = 3.2 \mathrm{nm}$ for spherical SiO$_2$ grains of $1.9 \mathrm{\mu m}$ in diameter. In this work we follow \citet{Paszun:2008} and set $\xi_\mathrm{crit} = 2 \mathrm{nm}$ for $1.2 \mathrm{\mu m}$ sized grains.

When a displacement exceeds its critical value it will be restored to the elastic limit. If for instance the rolling displacement exceeds its critical value $\|\vec{\xi}\| > \xi_\mathrm{crit}$, the contact pointers will be corrected $\vec{n}_1 \rightarrow \vec{n}_1^\mathrm{c}$, $\vec{n}_2 \rightarrow \vec{n}_2^\mathrm{c}$ and thus $\vec{\xi} \rightarrow \vec{\xi}^\mathrm{c}$ such that $\|\vec{\xi}^\mathrm{c}\| = \xi_\mathrm{crit}$. This modification of the contact pointers reflects a change of the corresponding potential energy. Therefore, we can keep track of the amount of dissipated energy. For further details on how the inelastic motion is applied we refer to \citet{Wada:2007}.

\begin{figure*}[th]
\resizebox{0.45\hsize}{!}{\includegraphics{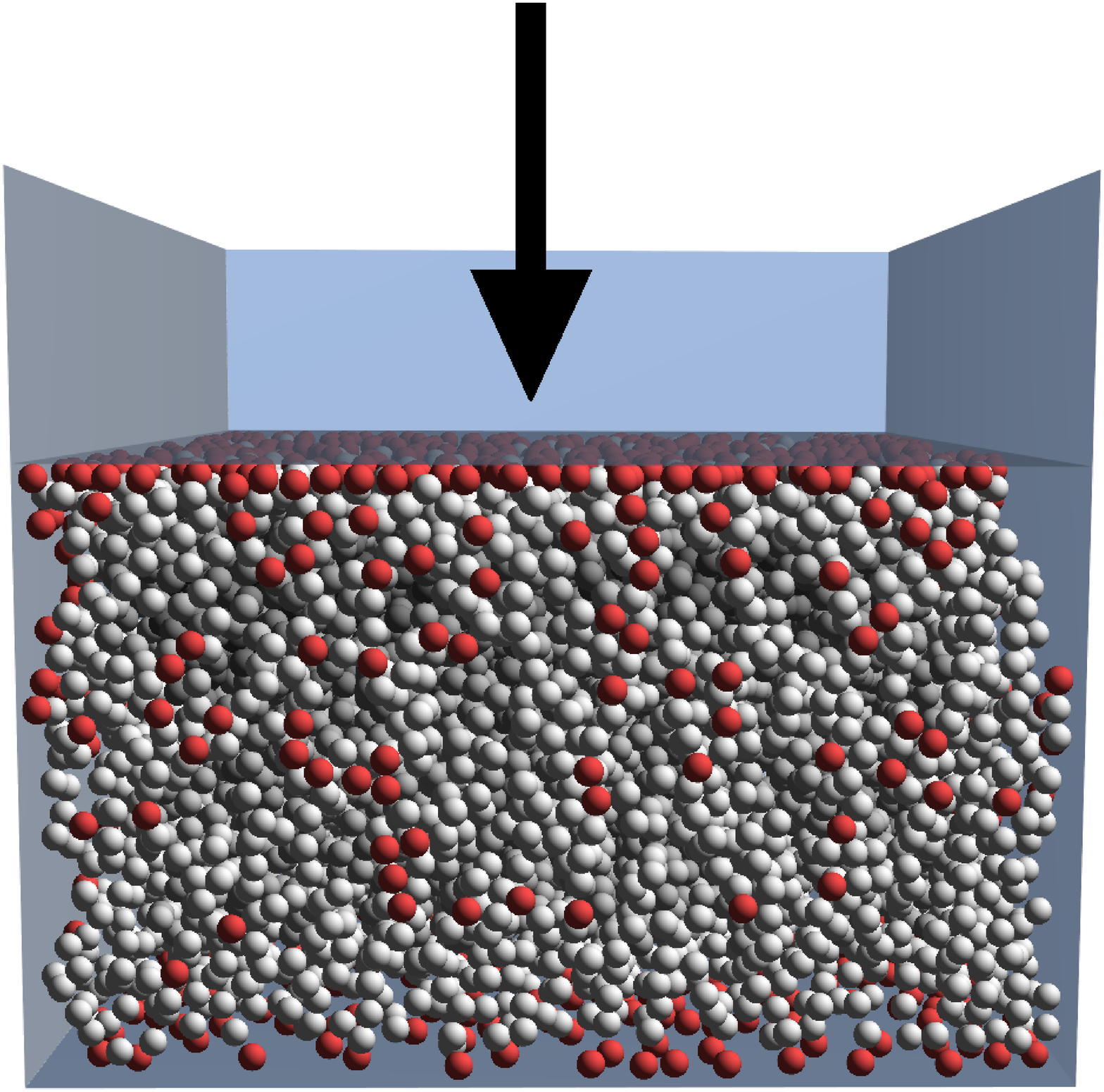}} \hfill \resizebox{0.45\hsize}{!}{\includegraphics{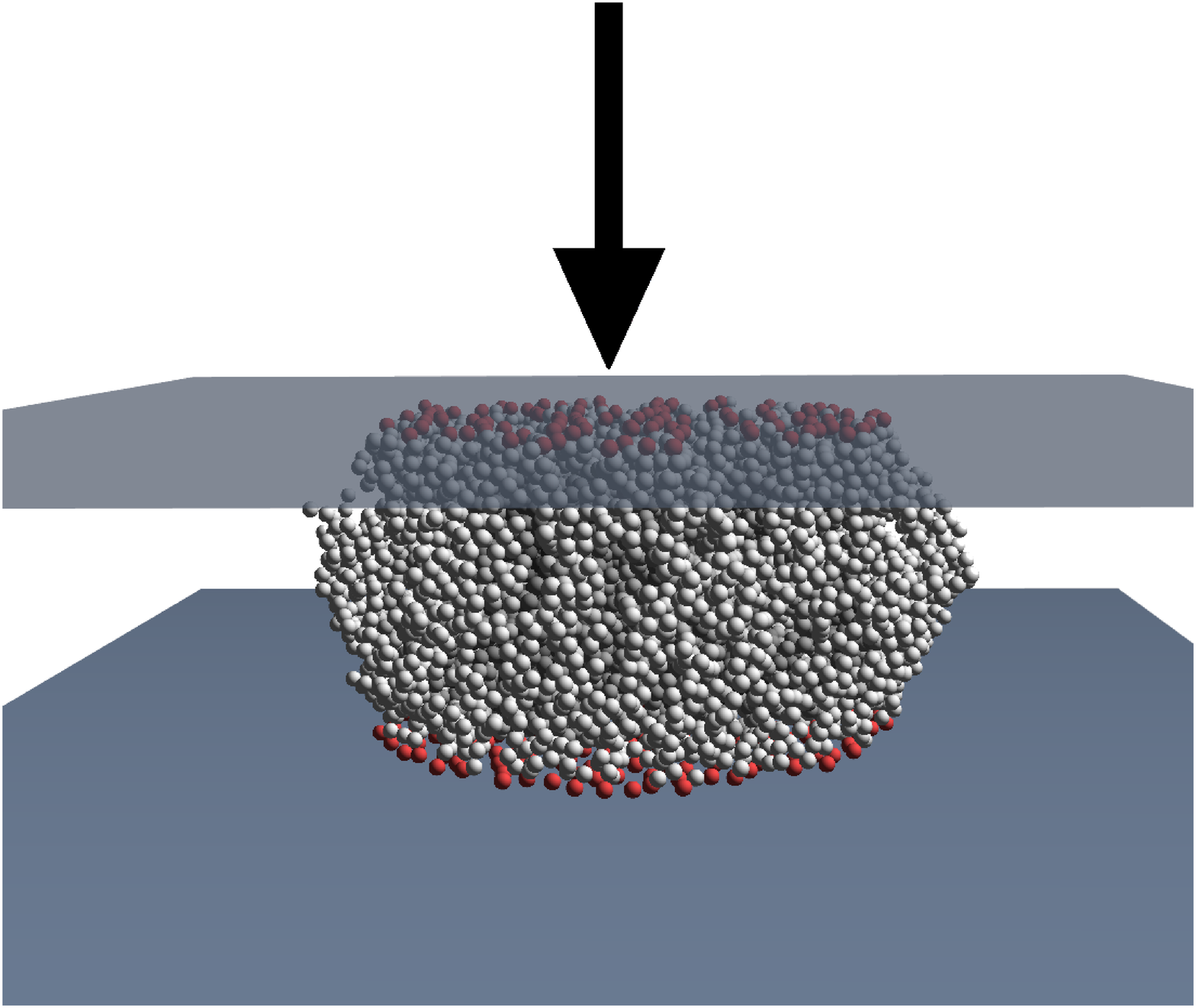}}
\caption{Setup of the numerical simulations to obtain material parameter for porous dust agglomerates. \textbf{Left:} Omnidirectional compression, while the top wall is moving downwards at constant speed, the sample is enclosed in a box with fixed walls on all sides.\ \textbf{Right:} Unidirectional compression, as the sample is getting compressed between two walls, particles can leave the initial volume to the sides. Particles colored red are actually in contact with the walls.}
\label{fig:cpr_setup}
\end{figure*}

\subsection{Particle-Wall interaction}
To study the compression of a sample aggregate it is necessary to constrain the motion of the monomers using suitable boundary conditions. In the corresponding experiments the sample has been put in a solid box with a movable piston on top \citep[see][Fig.\,2]{Guettler:2009}. In our simulations we model the experimental setup by putting the sample aggregate into a box of fixed walls. During the simulation the top wall is moving downwards with constant speed in order to compress the sample (see Fig.\,\ref{fig:cpr_setup}).

We assume that the monomers may interact with a wall in a similar way as they interact with other grains. In accordance with the particle-particle interaction we derive the corresponding forces and torques following the approach presented by \citet{Wada:2007}. For this purpose we assume that the wall can be described as a very huge particle in the limit $r_{\mathrm{wall}} \to \infty$.

\subsubsection{Compression}
The force of the particle-wall interaction in normal direction is very similar to the case of the particle-particle interaction. The compression length $\delta$ is given by
\[\delta = r - d\ ,\]
where $r$ is the radius of the monomer and $d$ denotes the distance between the surface of the wall and the center of the grain. Given a point $\vec{p}$ located on the surface of the wall and the surface normal $\vec{n}_w$,  we can easily obtain $d$ by
\[d = |\left( \vec{x} - \vec{p} \right) \cdot \vec{n}_w |\ ,\]
where $\vec{x}$  denotes the position of the particle. The force in normal direction can then be calculated using Eq.\,(\ref{F_normal}). Note that here the reduced radius $R$ is different to the case of particle-particle interaction. In the limit of $r_{\mathrm{wall}} \rightarrow \infty$ the reduced radius $R$ equals the radius $r$ of a monomer:
\[R = \lim_{r_{\mathrm{wall}} \to \infty} \frac{r\ r_{\mathrm{wall}}}{r + r_{\mathrm{wall}}} = r\ .\]

\subsubsection{Rolling}
\label{sec:inelastic_wall_rolling}
Keeping in mind that $R = r$, the torque $\vec{M}_\mathrm{r}$ acting on the particle caused by rolling along the surface of the wall is given by
\begin{eqnarray}\vec{M}_\mathrm{r} = k_\mathrm{r,wall} r^2 \vec{n}_1 \times \vec{n}_\mathrm{w},\label{eqn:M_roll_wall}\end{eqnarray}
where $k_\mathrm{r,wall}$ is equivalent to the rolling constant $k_\mathrm{r}$ given in Eq.\,(\ref{eqn:k_r}) taking the different reduced radius of the particle-wall interaction into account. $\vec{n}_\mathrm{w}$ denotes the surface normal of the wall. Note that it is important on which side of the wall the particle is located with respect to the direction of $\vec{n}_\mathrm{w}$. In the simulation we must either ensure that the particles remain on the positively oriented side all the time or check on which side of the wall the particle is located and correct the sign of Eq.\,(\ref{eqn:M_roll_wall}) if necessary.

To model the motion of a particle which is rolling inelastically over a wall we use a similar approach as for the inelastic particle-particle interaction. Taking $r_2 \rightarrow \infty$ into account, the correction of the contact pointers is then given by
\begin{eqnarray}
 \vec{n}_1^\mathrm{c} & = & \vec{n}_1 - \frac{\alpha}{r} \Delta \vec{\xi}\ ,\\
 \vec{n}_2^\mathrm{c} & = & \vec{n}_2 = \vec{n}_\mathrm{w}\ ,
\end{eqnarray}
where $\alpha$ is a correction factor derived in detail in \citet{Wada:2007} and $\Delta \vec{\xi}$ is given by
\begin{eqnarray}\Delta \vec{\xi} = \vec{\xi} \left(1 - \frac{\xi_\mathrm{crit}}{\|\vec{\xi}\|}\right)\ .\end{eqnarray}
The contact pointer $\vec{n}_2$ of the ''wall particle'' is equivalent to the normal vector $\vec{n}_\mathrm{w}$ of the wall and is not modified during the inelastic rolling motion. 

\subsubsection{Sliding}
To describe the sliding motion of a particle on a wall it is not suitable to start with Eq.\,(\ref{eqn:sliding_displacement}) and assume that $\vec{n}_\mathrm{c} \rightarrow  \vec{n}_w$ and $\vec{n}_2 \rightarrow \vec{n}_w$. Therefore we choose a different approach that takes into account how far the contact has slided from the location where it has initially formed. 

Let $\vec{p}$ denote the position, where the surface of the particle touched the wall when establishing the contact. For any later time, the center of the contact area is given by $\vec{x} + r \vec{n}_1$, where $\vec{x}$ is the current position of the particle. We define
\[\vec{\zeta}_0 = \vec{x} + r \vec{n}_1 - \vec{p}\ .\]
The sliding displacement is then given by
\begin{eqnarray}\vec{\zeta} = \vec{\zeta}_0 - \left(\vec{\zeta}_0 \cdot \vec{n}_\mathrm{w}\right) \vec{n}_\mathrm{w}\ .\end{eqnarray}

To model the inelastic wall sliding we modify the initial center of the contact area $\vec{p}$. If $\|\vec{\zeta}\| > \zeta_{\mathrm{crit}}$, we apply the correction
\begin{eqnarray}\vec{p}^\mathrm{c} = \vec{p} + \left(1 - \frac{\zeta_\mathrm{crit}}{\|\vec{\zeta}\|}\right) \vec{\zeta}\ .\end{eqnarray}

\subsubsection{Twisting}
The torque caused by the twisting motion is calculated in the same way as if two particles are in contact. Starting with the twisting displacement $\vec{\phi}$ given in equation (\ref{eqn:twist_displ}) we obtain
\begin{eqnarray} 
\vec{\phi}_\mathrm{wall} = \vec{n}_\mathrm{c} \int_{t_0}^t \left( \vec{\omega}_1(t')\right) \cdot \vec{n}_\mathrm{c}(t') dt'\ ,
\end{eqnarray}
under the assumption that the wall does not rotate. The torque is then given by
\begin{eqnarray} 
\vec{M}_\mathrm{t,wall} = - k_\mathrm{t,wall} \vec{\phi}_\mathrm{wall}\ ,
\end{eqnarray}
where $k_\mathrm{t,wall} = k_\mathrm{t}$. Here we assume that particles may twist elastically around the same angle for both the particle-particle and particle-wall interaction. According to test simulations where we measured the relative importance of the different wall interaction dissipation channels, inelastic wall twisting is of only minor importance.

%__________________________________________________________________
\section{Setup}
In order to calibrate the model and to compare simulations in detail with experimental results we focus here on two different numerical experiments that follow closely the experimental setup. Specifically, we will deal with agglomerates enclosed in a box and aggregates confined between two plates as depicted in Fig.\,\ref{fig:cpr_setup}.

\subsection{Sample generation}
In accordance with the mentioned laboratory experiments by \citet{Blum:2004,Guettler:2009}, our samples are produced by random ballistic deposition (RBD), which means that single grains are successively poured down on the existing aggregate impacting from the same direction. In order to prevent any restructuring upon impact the impact velocity of a monomer hitting the sample is kept very low. The resulting samples feature a filling factor $\phi$ between $0.12-0.15$. Filling factors of $0.12-0.14$ result from the fluffier surface and are therefore only observed for small samples. As the size of the samples increases this surface effect becomes negligible and the filling factor converges to $0.15$ which agrees well with the dust cakes used in the corresponding laboratory experiments and with numerical studies by \cite{Watson:1997}.

\subsection{Measurements}
\subsubsection{Pressure}
In the numerical (and laboratory) experiments a box-shaped sample is enclosed between walls that constrain the motion of the grains (see Fig.\,\ref{fig:cpr_setup}). Then, the top wall is being moved downwards at a constant speed until the filling factor exceeds a certain threshold $\phi_\mathrm{crit}$. Typically, $\phi_\mathrm{crit}$ is set to $0.7$.
As the top wall is moving downwards the volume of the sample decreases and it is increasingly more compressed. The total force $F_\mathrm{w}$ acting upon the top wall is calculated by summarizing the forces $\vec{F}_i$ exerted on the wall by grains that are currently in contact with it, where only the component in normal direction $\vec{n}_\mathrm{w}$ to the wall is taken into account
\[F_\mathrm{w} = \vec{n}_\mathrm{w} \cdot \left(\sum_i \vec{F}_i\right)\ .\]
The pressure $P$ is then given by
\[P = \frac{F_w}{A},\]
where $A$ denotes the base size of the box.

If there are only a few particles in contact with the wall, $F_w$ may change considerably from one integration step to the next due to the normal oscillations of the particle-wall contact (see Sect.\,\ref{sec:normal_damping}). Since these vibrations occur on a timescale of nanoseconds whereas the compression timescale is typically orders of magnitudes higher, it is reasonable to average over several integration steps (covering a few normal particle oscillations) to reduce the noise in the pressure determination. Typically, we averaged over 100 integration steps in this work.

\subsubsection{Filling factor}
The volume filling factor $\phi$ is defined as
\begin{eqnarray}\phi = \frac{V_\mathrm{mat}}{V_\mathrm{tot}},\end{eqnarray}
where $V_\mathrm{mat} = 4/3 \pi r^3 N$ denotes the volume of $N$ particles of radius $r$ and $V_\mathrm{tot}$ is the volume that the aggregate currently occupies. 
Calculating the filling factor is trivial in the case of omni-directional compression (see Fig.\,\ref{fig:cpr_setup}, left panel) as $V_\mathrm{tot} = A h$ for a box with base size $A$ and current height $h$. 

However, in the uni-directional case (Fig.\,\ref{fig:cpr_setup}, right panel) there are no side walls containing the aggregate, and particles will leave the initial volume. They flow to the sides as the top wall is moving downwards. This complicates the determination of $V_\mathrm{tot}$ in the numerical as well as experimental setup.
In the following we assume that the volume the aggregate is currently occupying is given by its projected cross section $A_\mathrm{proj}$ and the current height $h$ of the aggregate, which equals the distance between the top and the bottom wall. $A_\mathrm{proj}$ is obtained by projecting the aggregate in the plane of the top/bottom wall.

\subsection{Previous work}
A first attempt to determine the compressive strength numerically has been presented by \citet{Paszun:2008}. With the exception of the damping of the normal interaction (see Sects.\,\ref{sec:sticking_velocity} and \ref{sec:normal_damping}) they used the same particle interaction model that we use here. However, instead of flat walls they used two huge particles with radii much bigger than the dust grains to model the boundary conditions. The sample was put between the ``wall particles'' and the upper particle was being moved downwards at constant speed while the force acting on this ``wall particle'' was stored for later analysis. Since the motion of the grains has only been constrained in the vertical direction, grains could dodge to the sides as the pressure increased. Thus, an increase of the initial cross section of the aggregate was observed during the compression that led to a significant uncertainty in the determination of the volume the aggregate occupied at a certain point of time. Since the number of monomers was also limited to very small numbers ($\approx 300$) it is questionable if the results hold in the continuum limit.

\citet{Paszun:2008} confined themselves to the case of unidirectional compression. To our knowledge, the case of omnidirectional compression has not been simulated so far for this material. The compressive strength was only determined for a compression speed of $0.05\, \mathrm{m/s}$. A possible dependency of the compression behaviour on the speed of the compression has not been studied before.

%______________________________________________________________
\section{Numerical method}
To integrate the equations of motion we use a second order, symplectic Leap-Frog scheme. The main reason behind this choice lies in the fact that the forces and torques have to be determined only once during each integration step. Likewise, the evolution of the contact pointers and the twisting displacement (see Eq.\,(\ref{eqn:twist_displ})) is calculated with second order accuracy.

The timestep is limited by the oscillations in direction of the contact normal caused by the normal force (see Sect.\,\ref{sec:normal_damping}) as well as the oscillations in the tangential plane of the contact caused by the sliding force. As already mentioned the normal oscillation period is of the order of $10\,\mathrm{ns}$. 

The period of the oscillations in the tangential plane can be obtained in the following way: \citet{Wada:2007} derive the sliding force from the corresponding sliding potential $U_\mathrm{s}$
\begin{equation}U_\mathrm{s} = \frac{1}{2} k_\mathrm{s} \|\vec{\zeta}\|^2\ .\end{equation}
We can get an estimate of the oscillation period $T$ by
\begin{equation}T = \frac{2 \pi}{\omega} = 2 \pi \sqrt{\frac{m}{k_\mathrm{s}}}\ ,\end{equation}
where $\omega$ denotes the corresponding angular frequency of the oscillation and $m$ is the mass of a monomer. For the material parameters given in Tab.\,\ref{tab:material_parameters} we obtain $T = 12.8\,\mathrm{ns}$. 
Applying the sliding modifier $m_\mathrm{s} = 2.5$ (see Sect.\,\ref{sec:calibration}) the tangential oscillation period decreases to $T = 7.83\,\mathrm{ns}$, which limits our timestep to $0.3\,\mathrm{ns}$.

%______________________________________________________________
\section{Results}

\begin{figure}
\resizebox{\hsize}{!}{\includegraphics{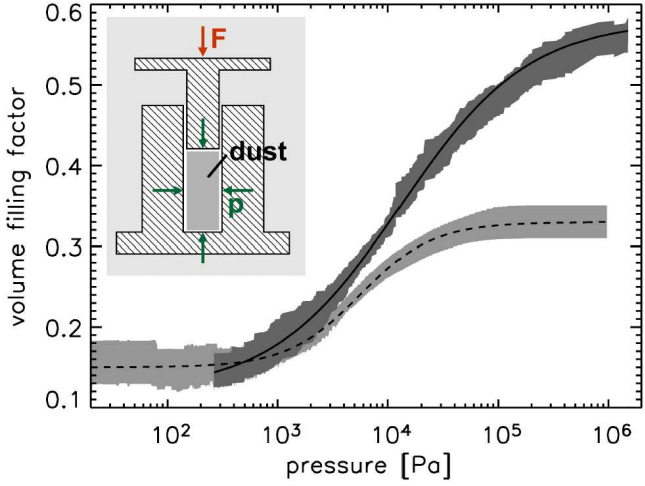}}
\caption{The compressive strength (filling factor $\phi$ versus pressure $p$) as obtained from experiments. The dark shaded region with the solid line fit refer to the omnidirectional compression experiment performed by \citet{Guettler:2009} whereas the light shaded region and the dashed line fit were obtained for unidirectional compression by \citet{Blum:2004}. The small image on the top left depicts the experimental setup of the omni-directional compression experiment. \citep[Figure taken from][]{Guettler:2009}}
\label{fig:cpr_exp}
\end{figure}

Calibration experiments using a continuum SPH-code indicate that the compressive strength of a porous dust aggregate depends on how fast the compaction takes place \citep{Guettler:2009, Geretshauser:2010}. So far, in laboratory experiments the compressive strength could only be determined for a slow quasi-static compression process,
where the compressed aggregate had been given sufficient time for relaxation (in the following referred to as static compression) \citep{Guettler:2009}. 

The static compression provides us with the possibility to check how well our model is able to describe the compression behaviour of porous dust aggregates. In the first step we will therefore use the case of the omnidirectional static compression to calibrate our molecular dynamics model. Afterwards we will increase the speed of the top wall. At a sufficiently high speed the relaxation of the aggregate will not be possible any longer (from now on referred to as dynamic compression).

\subsection{Calibration of our model}
\label{sec:calibration}
To model the quasi-static compression the speed $v_\mathrm{wall}$ at which the top wall is moving downwards should be as low as possible. As the time required to reach a certain filling factor is inversely proportional to $v_\mathrm{wall}$, the runtime of the simulation constitutes a lower limit of $v_\mathrm{wall}$. \citet{Paszun:2008} considered $v_\mathrm{wall} = 5\, \mathrm{cm/s}$ to be low enough to model static compression. However, we observe that the curves still change when using even lower velocities (see Fig.\,\ref{fig:omnidirectional_compression_velocity_low} below). To model the case of static compression we set $v_\mathrm{wall} = 1\, \mathrm{cm/s}$. To ensure it is a reasonable choice we checked lower velocities down to $v_\mathrm{wall} = 0.2\, \mathrm{m/s}$ but observed only a tiny deviation of the resulting curves.

Depending on the number of particles we use, the diameter of our sample aggregates is roughly $40-60\, \mathrm{\mu m}$ which is about $\times 10^3$ times smaller than the samples used in the laboratory experiments. While the sample is getting compacted the particles on the edges of the sample must overcome the sliding resistance of the side wall. Owing to the small diameter of the sample this has a significant influence on the resulting force on the top wall. To mitigate this effect we reduce the strength of the rolling, sliding, and twisting interaction between particles and the side walls by a factor of $1000$.

\begin{figure}
\resizebox{\hsize}{!}{\includegraphics{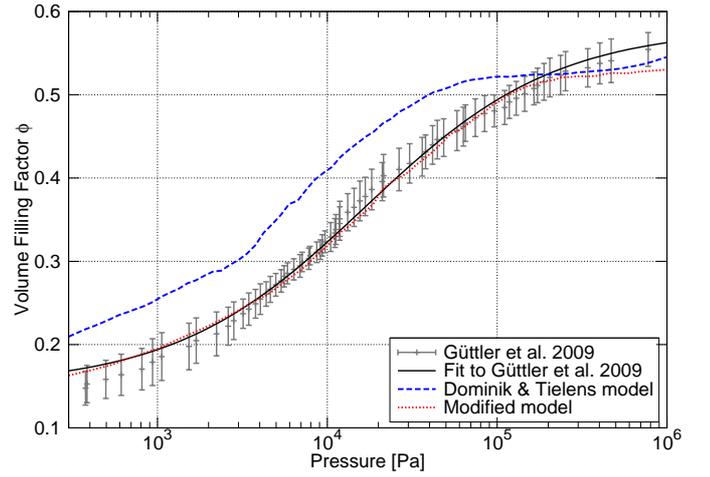}}
\caption{Compressive strength for omnidirectional compression of cubical aggregates. The result of the unaltered interaction model is compared to our improved model. The black line represents a fit to experimental data obtained from \citet{Guettler:2009}.}
\label{fig:omnidirectional_compression}
\end{figure}

The results of the corresponding laboratory experiment are shown in Fig.\,\ref{fig:cpr_exp}. The solid black curve is a good fit to the omnidirectional experimental data (see Fig.\,\ref{fig:omnidirectional_compression} and dark shaded area in Fig.\,\ref{fig:cpr_exp}) and is given by
\begin{eqnarray}
  \phi(P) = \phi_2 - \frac{\phi_2 - \phi_1}{\exp \left( \frac{\log_{10} P - \log_{10} p_\mathrm{m}}{\Delta} \right) + 1}\ .
  \label{eqn:compressive_strength}
\end{eqnarray}
Using $\phi_1 = 0.15$ and $\phi_2 =  0.58$ we obtain $p_\mathrm{m} = 16.667\,\mathrm{kPa}$ and $\Delta = 0.562$. The black curve shown in Fig.\,\ref{fig:omnidirectional_compression} depicts this fit.

The results of the unidirectional experiments, also displayed in Fig.\,\ref{fig:cpr_exp}, have been fitted by \cite{Blum:2004} using a similar curve with $\phi_1 = 0.15$, $\phi_2 = 0.33$, $p_m = 5.6\,\mathrm{kPa}$, and $\Delta = 0.33$.

\subsection{Omnidirectional compression}
\label{sec:omnidirectional_compression}

\subsubsection{Quasi-static case}

In Fig.\,\ref{fig:omnidirectional_compression}, we compare the experimental fitting curve to results from our simulations of box-shaped samples featuring approximately $11000$ particles and an edge length of about $40\, \mathrm{\mu m}$. All curves have been obtained from averaging the results of six independent runs with samples having statistically equal bulk properties. 

Obviously, simulations using the Dominik and Tielens model do not reproduce the experimental data very well. For low pressures the blue dashed curve in Fig.\,\ref{fig:omnidirectional_compression} lies significantly above the solid black one, i.e.\ applying the same pressure the sample has been compressed to a higher filling factor in the simulations than compared to the experiments. In order to make it harder to compress an aggregate we tried to increase the stiffness of the aggregates by modifying the particle interaction.

To our knowledge the equations describing the rolling and sliding interaction have not been experimentally tested yet, whereas the pull-off force of the normal interaction has been measured using atomic force microscopy \citep{Heim:1999}. Thus, we vary the strength of the rolling and sliding interaction. For this purpose we simply multiply the constants $k_\mathrm{r}$ and $k_\mathrm{s}$ (see Eqns.\,\ref{eqn:k_r} and \ref{eqn:k_s}) with correction factors that we further refer to as rolling/sliding modifiers $m_\mathrm{r}$ and $m_\mathrm{s}$. This constitutes a straightforward approach to increase the stiffness of monomer chains. In fact, we also modified the strength of the twisting interaction but found that it had very little to no impact on the compressive curve. Therefore we do not alter twisting in this work.

Choosing $m_\mathrm{r} = 8$ and $m_\mathrm{s} = 2.5$, we obtain the red-dotted curve in Fig.\,\ref{fig:omnidirectional_compression}. All in all, our modified interaction model is able to reproduce experimental results much better than the original version. In particular, for pressures below $100\,\mathrm{kPa}$ we observe a very good agreement with experimental data. However, we observe a deviation for pressures above $300\,\mathrm{kPa}$. In our simulations a pressure of more than $1\,\mathrm{MPa}$ is required to further compress aggregates when a filling factor of about $\Phi = 0.52$ is reached.

\begin{figure}
\resizebox{\hsize}{!}{\includegraphics{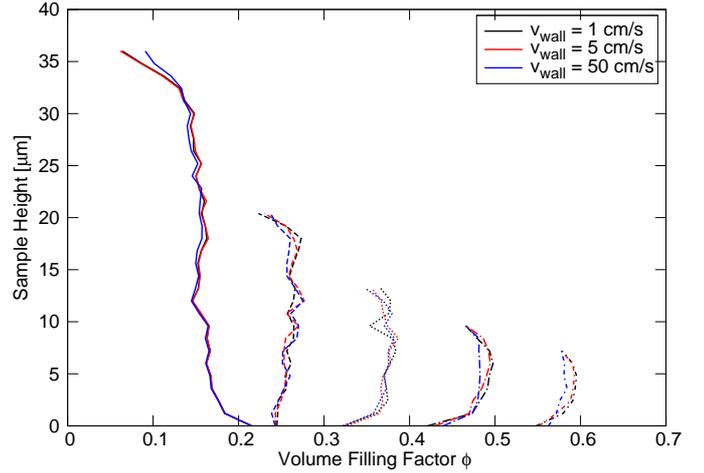}}
\caption{Snapshots of the vertical profile of the aggregate's filling factor averaged horizontally. Shown are results at 5 evolutionary times using 3 different speeds. The left (solid) curves represent the initial state at $t=0$. As the top wall moves slowly downwards the aggregate is compacted almost homogeneously as is indicated by the nearly vertical curves. At the top and bottom of the box wall effects produce a slight deviation.}
\label{fig:filling_factor_profile_static}
\end{figure}

In the quasi-static case the aggregate is given sufficient time to restructure and counteract the pressure exerted on the walls. Thus, we expect the filling factor increases homogeneously in the sample. In Fig.\,\ref{fig:filling_factor_profile_static} the vertical profile of the filling factor is plotted for different stages of the compression process. To determine the filling factor profile, the sample is split vertically into equidistant intervals with the length of one particle diameter. Then, the average filling factor is calculated for each interval. Note that the fractal structure resulting in a filling factor of $\phi = 0.15$ in the bulk part of RBD-generated aggregates is not present at the bottom of the aggregate. Therefore the filling factor there exceeds the average value of $\phi = 0.15$.

During the compression process several snapshots of the filling factor profile have been taken. As expected the filling factor increases almost homogeneously for slow compression speeds. Keep in mind that it requires higher pressure to compact particles close to a wall since neighbouring particles have to be pushed away. Therefore the filling factor of the particle layers close to the top or bottom wall is lower compared to the rest of the aggregate for highly compacted aggregates. This effect causes the crescent shaped curves observed for highly compacted aggregates in Fig.\,\ref{fig:filling_factor_profile_static}.

\begin{figure}
\resizebox{\hsize}{!}{\includegraphics{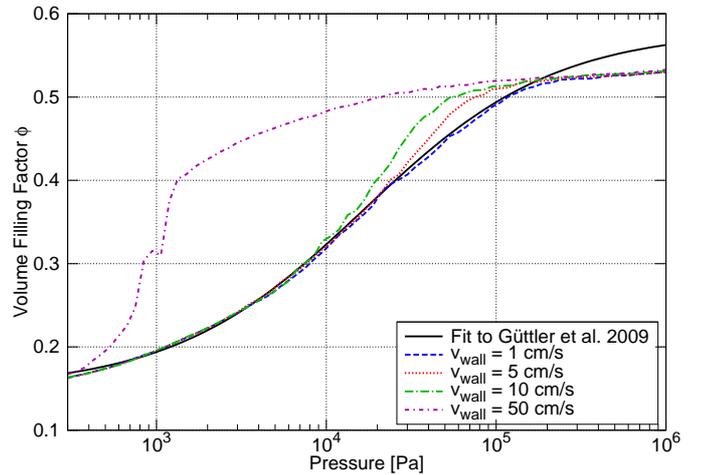}}
\caption{Compressive strength for low velocity omnidirectional compression.}
\label{fig:omnidirectional_compression_velocity_low}
\end{figure}

The corresponding compressive strength curve for these low velocities is shown in Fig.\,\ref{fig:omnidirectional_compression_velocity_low}. For compression speeds of $5\,\mathrm{cm/s}$ and below the differences to the quasi-static case remain small. For higher velocities we observe an increasing deviation from the quasi-static curve. Thus, the results from \citet{Guettler:2009} cannot be applied for higher velocities.

\subsubsection{Dynamic case}

\begin{figure}
\resizebox{\hsize}{!}{\includegraphics{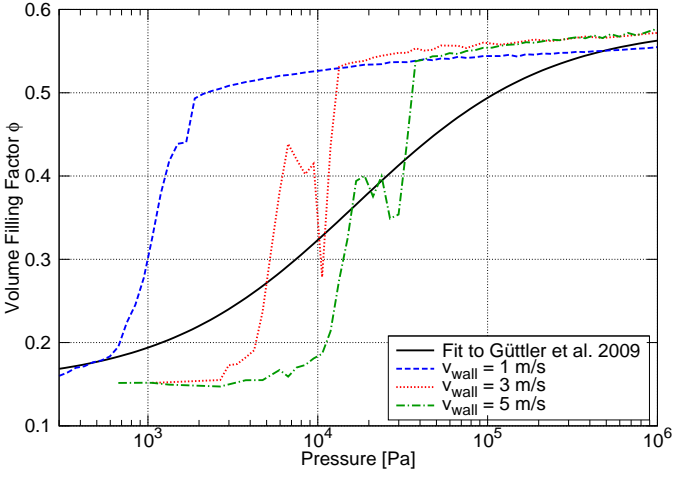}}
\caption{Dynamic omnidirectional compression with $v_\mathrm{wall}$ between $1-5 \mathrm{m/s}$.}
\label{fig:omnidirectional_compression_velocity_high}
\end{figure}

\begin{figure}
\resizebox{\hsize}{!}{\includegraphics{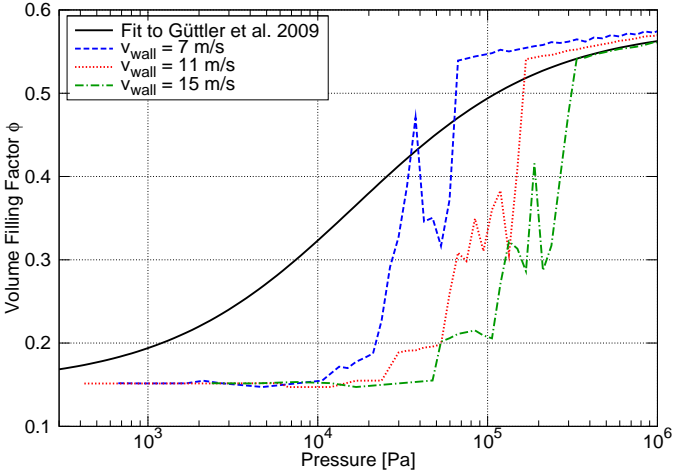}}
\caption{Dynamic omnidirectional compression with $v_\mathrm{wall}$ between $7-15 \mathrm{m/s}$.}
\label{fig:omnidirectional_compression_velocity_high2}
\end{figure}

In a second step, we studied the influence of large compression speeds. 
If the compression speed exceeds around $1\, \mathrm{m/s}$ the compression behaviour changes considerably. The required simulation time is inversely proportional to the compression speed. Therefore we use a higher number of particles for compression experiments with wall speeds above $1\, \mathrm{m/s}$. The box-shaped samples are composed of about $40000$ particles and feature a base length and height of $\approx 60\,\mathrm{\mu m}$. Compared to the quasi-static case the shape of the curves changes drastically, see Figs.\,\ref{fig:omnidirectional_compression_velocity_high} and \ref{fig:omnidirectional_compression_velocity_high2}. Instead of a smooth transition, three distinguished regimes emerge: The filling factor does not increase until a certain critical pressure is reached. Then, only a small additional pressure is required to compact the aggregate. When the aggregate is close to its final compaction the pressure again increases sharply.

\begin{figure}
\resizebox{\hsize}{!}{\includegraphics{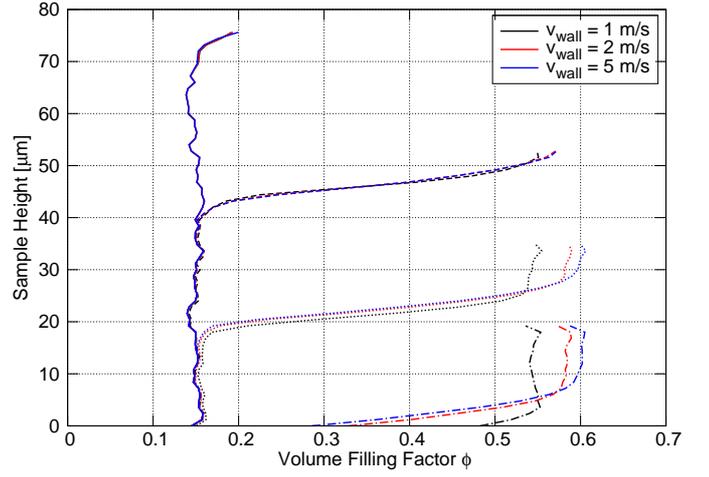}}
\caption{Snapshots of the vertical profile of the filling factor. As the top wall moves downwards rapidly the compaction of the lower parts of the aggregate is lagging behind. The color indicates the compression speed whereas the line type indicates the position of the top wall. The solid lines show the filling factor profile of the initial uncompressed sample.}
\label{fig:filling_factor_profile_dynamic}
\end{figure}

This can be easily explained by looking at Fig. \ref{fig:filling_factor_profile_dynamic}. When the compression speed exceeds a value of $1\, \mathrm{m/s}$ the aggregate is compacted inhomogeneously. The compression occurs too fast to allow the propagation of the top pressure through the entire sample. We clearly see the emergence of a very dense layer right beneath the top wall. This compact layer propagates downwards at the speed of the wall similar to a snowplough clearing freshly fallen snow. While pushing the dense layer downwards the pressure remains constant. After it reaches the bottom of the sample the pressure required to compress the sample a little bit further increases drastically. The sharp spikes shown in Figs.\,\ref{fig:omnidirectional_compression_velocity_high} and \ref{fig:omnidirectional_compression_velocity_high2} result from the highly compacted layer reaching the bottom of the sample. The density wave is reflected from the bottom causing heavy fluctuations of the pressure on the top and bottom wall.

By comparing the filling factor profile during the compression to the initial one we can determine at which speed $v_{\mathrm{p}}$ the compaction is propagating downwards trough the sample. For this purpose we measure the height where the initial and current filling factor profile deviate from each other and use the time that has passed since the start of the simulation to calculate the speed. Averaging over six different samples we obtain $v_{\mathrm{p}} = 6.98 \pm 0.16\, \mathrm{m/s}$.

To provide continuum-simulations with a simple recipe for the dynamic compressive strength we performed simulations using compression speeds up to $25\,\mathrm{m/s}$. A few examples are shown in Figs.\,\ref{fig:omnidirectional_compression_velocity_high} and \ref{fig:omnidirectional_compression_velocity_high2}. For every compression speed we determine a fit curve similar to Eq.\,(\ref{eqn:compressive_strength}), where $p_m$ and $\Delta$ serve as fitting parameters. Thus, we obtain values of $p_m$ and $\Delta$ for different compression speeds. In the last step we determine for each parameter an analytic approximation describing the dependency on the compression speed.

We observe different behaviour between the the quasi-static case for low velocities and the dynamic case for higher velocities. In the beginning, $p_m$ decreases which means that the sample can be compressed more easily. This can be explained by the fact, that the aggregate is given less time to restructure and counteract the external pressure exerted on it by the wall. However, this effect will be reversed when the compression speed exceeds a critical value of $v_{\mathrm{crit}} \approx 0.9\,\mathrm{m/s}$. In the dynamic regime, it gets considerably harder to compress the sample with increasing velocity of the wall. Therefore it is helpful to distinguish between the two regimes.

\begin{figure}
\resizebox{\hsize}{!}{\includegraphics{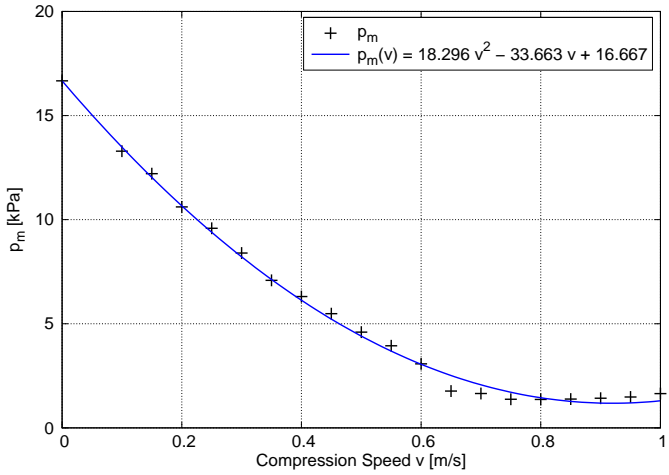}}
\caption{Dependence of the fit parameter $p_m$ of Eq.\,(\ref{eqn:compressive_strength}) on the compression speed $v$ in the quasi static regime of $v \leq 1\,\mathrm{m/s}$.}
\label{fig:fit_pm_static}
\end{figure}

In Fig.\,\ref{fig:fit_pm_static} the dependence of $p_m$ on the compression speed $v$ is shown for values of $v \leq 1\,\mathrm{m/s}$. Using the ansatz $p_m(v) = a v^2 + b v + c$ we obtain
\begin{equation} p_m(v) = \left( 18.296\,v^2 - 33.663\,v + 16.667\right)\,\mathrm{kPa}\ ,\label{eq:fit_pm_static}\end{equation}
where the compression speed $v$ is given in m/s. 

\begin{figure}
\resizebox{\hsize}{!}{\includegraphics{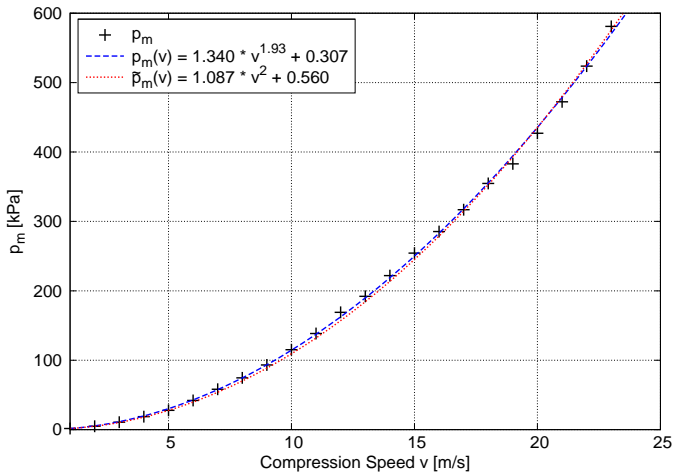}}
\caption{Dependence of the fit parameter $p_m$ of Eq.\,(\ref{eqn:compressive_strength}) on the compression speed $v$ in the dynamic regime of $v \geq 1\,\mathrm{m/s}$.}
\label{fig:fit_pm_dynamic}
\end{figure}

In Fig.\,\ref{fig:fit_pm_dynamic} the dependence of $p_m$ on the compression speed $v$ is shown for values of $v \geq 1\,\mathrm{m/s}$. To find a simple analytic approximation we choose a power law of the form $p_m(v) = a v^b + c$ and we obtain
\begin{equation} p_m(v) = ( 1.340 \, v^{1.93} + 0.307) \, \mathrm{kPa}\ .\label{eq:fit_pm_dynamic}\end{equation}
The fitted parabola describes the data points well. As the exponent of $1.93$ is close to $2$, we determine a second fit where the exponent was set to $2$ and get
\begin{equation} \tilde{p}_m(v) = ( 1.087 \, v^{2} + 0.560) \, \mathrm{kPa}\ .\label{eq:fit_pm_dynamic2}\end{equation}

\begin{figure}
\resizebox{\hsize}{!}{\includegraphics{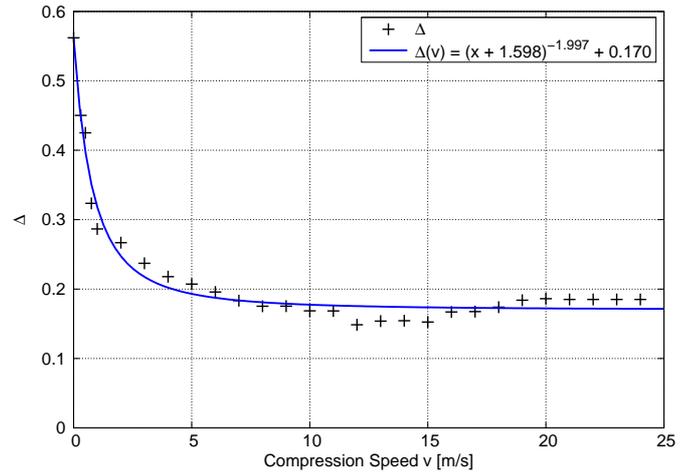}}
\caption{Dependence of the fit parameter $\Delta$ of Eq.\,(\ref{eqn:compressive_strength}) on the compression speed.}
\label{fig:fit_delta}
\end{figure}

Similarly, for the parameter $\Delta$ (see Fig.\,\ref{fig:fit_delta}) we obtain
\begin{equation} \Delta(v) = (v + 1.598)^{-1.997} + 0.170 \label{eq:fit_delta}\ .\end{equation}

\subsection{Unidirectional compression}
\begin{figure}
\resizebox{\hsize}{!}{\includegraphics{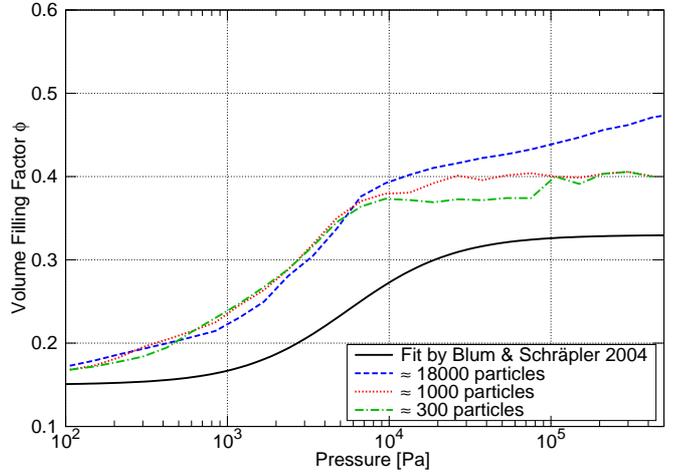}}
\caption{Compressive strength for unidirectional compression of cylindrical aggregates of different size using the original Dominik \& Thielens model.
  The black line has been obtained from experiments by \citet{Blum:2004}.}
\label{fig:unidirectional_compression}
\end{figure}

Additionally, we simulated the unidirectional compression of cylindrical samples of different sizes using the non modified model. The results are shown in Fig.\,\ref{fig:unidirectional_compression} where again each curve is obtained by averaging the results for six different samples of equal size. To compare our results to \citet{Paszun:2008} the speed of the top wall was set to $v_\mathrm{wall} = 5\, \mathrm{cm/s}$. Apparently there is a noticeable discrepancy between our simulations and the laboratory results. As in the case of omnidirectional compression, the pressure required to reach a certain filling factor is significantly lower in our simulations.

As we can see in Fig.\,\ref{fig:unidirectional_compression} the deviation for pressures above $10\,\mathrm{kPa}$ becomes more apparent if we increase the size of the samples. As \citet{Paszun:2008} compressed very small samples using only about 300 particles this may be the reason why their results showed better agreement with laboratory results for higher pressures. However, their compressive strength curve was also shifted in the same direction as in this work.

\begin{figure}
\resizebox{\hsize}{!}{\includegraphics{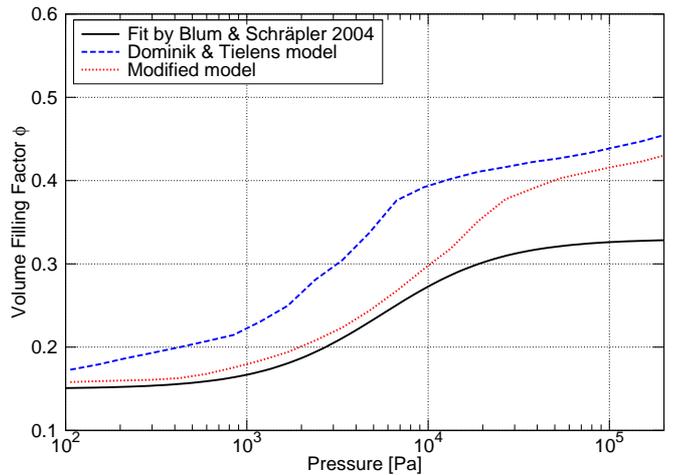}}
\caption{Unidirectional compression of cylindrical aggregates using the modified model with 18,000 particles. The black line has been obtained from experiments by \citet{Blum:2004}.}
\label{fig:unidirectional_compression_mod}
\end{figure}

Afterwards we tested the modified model with the same $m_\mathrm{r} = 8$ and $m_\mathrm{s} = 2.5$ as found above. As it is shown in Fig.\,\ref{fig:unidirectional_compression_mod} (red dotted curve), the modified model agrees very well with the experimental results for pressures below $10^4\, \mathrm{Pa}$. However, we still end up with considerably higher filling factors. A possible explanation is given by the fact that the size of the samples used in the laboratory experiments was of the order of centimeters and thus about 2000 times larger than our samples. \citet{Blum:2004} measured that the projected cross section increased by a factor of $1.6$ during the compression process. To reach similar pressures in our simulations the sample has to be compressed until its height is only about two times the diameter of a single monomer where the cross section increased roughly by a factor of $4$. Due to the larger diameter of the laboratory samples it is harder for monomers in the center to flow in the outward direction.

%______________________________________________________________
\section{Conclusions}
We have performed molecular dynamics simulations to study the compressive strength of dust agglomerates which plays an important role in determining the outcome of mutual collisions. Using a special setup for the simulations we were able to compare our results in detail to the outcome of of dedicated laboratory experiments.

Our simulations using the frequently applied interaction model by \citet{Dominik:1997} indicate that real aggregates composed of micron sized silicate grains feature a greater stiffness. Since the primary bulk properties of material used for the individual monomers are known experimentally very well (see Table~1), we decided to vary the force constants ($k_r$ and $k_s$) for rolling and sliding. Indeed, the higher stiffness can be accommodated by an increase of $m_r=8$ and $m_s=2.5$ (for $k_r$ and $k_s$, respectively) in comparison to the quoted values in Eqns.\,(\ref{eqn:k_r}) and (\ref{eqn:k_s}). After modifying the interaction model as described in Sect.\,\ref{sec:calibration}, we have been able to reproduce the experimental results much better, and found very good agreement for both, unidirectional and omnidirectional compression. This work reveals the importance of the rolling and sliding interaction for the restructuring of aggregates. As these interactions currently lack experimental testing we feel it desirable to study in particular the rolling and sliding of micron sized grains in laboratory experiments.

We have also studied the influence of the wall speed on the compression behaviour. If an aggregate is compressed slowly the filling factor increases homogeneously and the pressure needed to further compact the aggregate increases with increasing filling factor. For higher compression velocities a compacted layer emerges underneath the moving wall, similar to the shovel of a snow plow when pushing away snow. Once this layer has formed the pressure remains nearly constant until the layer has reached the bottom of the sample. The transition from the static towards the dynamic case occurs at compression speeds of the order of $1\,\mathrm{m/s}$. Since impact velocities of typical collisions of planetesimals lie within the range of $1-10\, \mathrm{m/s}$ the dynamic compression behaviour must be taken into account when simulating such collisions. 

To determine the impact of the rescaling of the rolling and sliding forces on the very early phases in the planetesimal formation process, we plan to perform detailed collision simulations for a wide set of collision parameter. This will allow us to calculate ab-initio the division between bouncing, sticking and fragmentation. This has recently been under experimental and theoretical scrutiny \citep{Zsom:2010, 2011A&A...531A.166G}. It is hard to estimate the consequences of the new (stiffer) parameter set on the outcome of agglomerate collisions. We suspect that this can possibly lead to an enhanced sticking as more energy can be stored in the system before it breaks. Using a much larger particle number than previously will allow us to determine more accurately important bulk parameter such as the sound speed.

\begin{acknowledgements}
A. Seizinger acknowledges the support through the German Research Foundation (DFG) grant KL 650/16.
Very helpful discussions with Carsten G\"uttler and J\"urgen Blum from the experimental laboratory astrophysics group at Braunschweig, as well as with Carsten Dominik, Hidekazu Tanaka and Ralf Geretshauser are gratefully acknowledged.
The authors acknowledge support through DFG grant KL 650/7 within the collaborative research group FOR\ 759 {\it The formation of planets}.
\end{acknowledgements}

\bibliographystyle{aa}
\bibliography{references}

\end{document}